\documentclass[11pt, table]{article}

\usepackage[english]{babel}
\usepackage[utf8]{inputenc}
\usepackage[T1]{fontenc}
\usepackage{lmodern}

\usepackage{natbib}
\bibpunct{(}{)}{,}{a}{,}{,}

\usepackage{amsmath, amssymb, amsfonts, amsthm}
\usepackage{cancel}
\usepackage{mathrsfs}
\usepackage{dsfont}

\numberwithin{equation}{section}

\usepackage{graphicx}
\usepackage[format=hang, font=small, labelfont=bf]{caption}
\usepackage{tikz}
\usetikzlibrary{decorations.pathreplacing}
\usetikzlibrary{positioning}

\usepackage{array}
\usepackage{multirow, multicol}
\usepackage{color}
\definecolor{gris85}{gray}{0.85}
\usepackage[algo2e,ruled]{algorithm2e}
\usepackage{setspace}

\usepackage{enumitem}

\usepackage[a4paper,citecolor=blue,linkcolor=black,colorlinks=true, urlcolor=black,filecolor=black]{hyperref}

\usepackage{authblk} 

\usepackage{geometry}
\usepackage{fancyhdr}
\renewcommand{\mid}{\;\ifnum\currentgrouptype=16 \middle\fi|\;}

\newcommand*{\esp}{\mathbf{E}}
\newcommand*{\ind}{\mathbf{1}}

\newcommand*{\grad}{\mathbf{\nabla}}

\newcommand*{\dd}{\mathrm{d}}

\newcommand*{\zero}[1][d]{\mathbf{0}_{#1}}

\usepackage{xspace}

\newcommand*{\eg}{\textit{e.g.}\xspace}

\newcommand*{\s}{\mathscr{S}}
\newcommand*{\G}{\mathscr{G}}
\newcommand*{\ivj}{i\sim j}
\newcommand*{\N}{\mathscr{N}}

\newcommand*{\bX}{\mathbf{X}}
\newcommand*{\bx}{\mathbf{x}}
\newcommand*{\Xset}{\mathscr{X}}

\newcommand*{\y}{\mathbf{y}}

\newcommand*{\bU}{\mathbf{U}}
\newcommand*{\bu}{\mathbf{u}}

\newcommand*{\btheta}{\boldsymbol\theta}
\newcommand*{\bTheta}{\boldsymbol\Theta}

\newcommand*{\aux}{\mathbf{r}}
\newcommand*{\step}{\epsilon}
\newcommand*{\mass}{\mathbf{M}}
\newcommand*{\pot}{\mathrm{V}}
\newcommand*{\gradE}[2]{\widehat{\grad}_{\bu^{(1,#2)},\ldots,\bu^{(#1,#2)}}}

\begin{document}
\title{\textsc{Noisy Hamiltonian Monte Carlo for doubly-intractable distributions}}
\author{\textsc{Julien Stoehr, Alan Benson, Nial Friel}}
\affil{School of Mathematics and Statistics, University College Dublin and Insight Centre for Data Analytics}
\date{}
\maketitle

\begin{abstract}
Hamiltonian Monte Carlo (HMC) has been progressively incorporated within the statistician's toolbox as an alternative sampling method in settings when standard Metropolis-Hastings is inefficient. HMC generates a Markov chain on an augmented state space with transitions based on a deterministic differential flow derived from Hamiltonian mechanics.  In practice, the evolution of Hamiltonian systems cannot be solved analytically, requiring 
numerical integration schemes. Under numerical integration, the resulting approximate solution no longer preserves the measure of the target distribution, therefore an accept-reject step is used to correct the bias. For doubly-intractable distributions -- such as posterior distributions based on Gibbs random fields -- HMC suffers from some computational difficulties: computation of gradients in the differential flow and computation of the accept-reject proposals poses difficulty. In this paper, we study the behaviour of HMC when these quantities are replaced by Monte Carlo estimates.
\end{abstract}

\noindent%

\section{Introduction}

Developing satisfactory methodology for the Bayesian analysis of statistical models with intractable likelihood functions is of considerable interest. Such models are motivated by a wide range of applications including 
spatial statistics, social network analysis, population genetics and image analysis. 
The challenges raised by such models stem from mathematical reasons -- the likelihood function does not admit a closed form as a function of $\btheta$ -- or computational reasons -- the likelihood function evaluation 
is time consuming. Such issues appear in settings where the likelihood function is known up to a parameter dependent normalising constant, that is, 
\begin{align}
\frac{
1
}{
Z(\btheta)
}
q(\btheta,\bx), 
\quad\text{where}\quad 
Z(\btheta) = \int_{\Xset} q(\btheta,\bx)\mu(\dd\bx),
\label{eq:likefunc}
\end{align}
which is the focus of this paper. One such class of models are Markov random fields (MRFs), undirected graphical models such as the Ising or Potts model which are used in a wide range of applications to model the 
dependency structure of correlated data.  Applications may be found in epidemiology \citep{green2002}, genetic analysis \citep{francois2006}, ecology \citep{augustin1998}, image analysis \citep[\eg][]{hurn2003}, 
amongst others. In social network analysis, the exponential random graph model (ERGM) \citep{robins2007} can be used to model the structure of a social network represented using a directed or undirected graph where 
edges in the graph show connections between the nodes, \eg, friendship. For ERGMs, the likelihood function \eqref{eq:likefunc} is constructed over binary adjacency matrices $\bx \in \Xset$ which represent the graph and 
the likelihood function is intractable since the number of graphs in a network with $n$ actors grows as $2^{\binom{n}{2}}$ for the undirected edge networks and $2^{n(n-1)}$ in the directed edge networks. 

Dealing with intractable likelihood functions has led to the development of important theoretical and methodological advancements in Bayesian statistics. A first approach to overcome the model's intractable bottleneck 
is to replace the true model with a pseudo-model selected among a collection of much simpler and more tractable set of probability distributions (e.g.\ variational Bayes \citep{jaakkola2000}, mean-fields approximation 
for MRFs \citep{jordan1999} or with an easily-normalised full conditional distributions (e.g.\ composite likelihood \citep{lindsay1988} and pseudolikelihood for MRFs \citep{besag1974}). However, these cruder 
approximations to the true model often miss some of the features of the original intractable model and can lead to unreliable estimates of the model parameters \citep{friel2012, stoehr-friel2015}.

Another point of view arises from sampling methods. Markov Chain Monte Carlo (MCMC) methods are arguably the most popular methodology, but Approximate Bayesian Computation \citep[ABC,][]{marin2012}, another 
simulation based approach has recently generated much activity in the literature. ABC deals with the situation where the likelihood function cannot be evaluated, due to intractability or otherwise, but can be 
simulated from. When performing parameter estimation, the method is particularly well suited for problems where the likelihood function does not admit an algebraic form, a situation where MCMC methods are at a 
loss but which we don't explore here. MCMC methods produce ergodic estimates of functionals with respect to the posterior of interest such as the expectation of parameters and so on. While some Bayesian 
estimators can be efficiently 
estimated with such methods via the empirical distribution, most methods cannot be used with a parameter dependent intractable likelihood function. Indeed, to produce a Markov chain which is reversible with respect 
to the posterior distribution, 
the method performs at each iteration an accept-reject step which requires computation of a ratio of intractable normalising constants. This 
problem is sometimes referred to as doubly-intractable Bayesian inference. \cite{murray2006}, extending the work of \cite{moller2006}, derive MCMC methods to handle the presence of intractable normalising terms in 
the acceptance ratio. The target distribution can be estimated without bias by using auxiliary variables whose proposal distributions have the relevant normalising constant. The latter solution leads to replacing 
the acceptance probability of the accept-reject step by a single point importance sampling estimate. Such ideas have appeared in the generalised importance Metropolis-Hastings of \cite{beaumont2003} and have later 
been extended by \cite{andrieu2009}.

For well-known MCMC methods, such as the Metropolis-Hastings algorithm \citep{metropolis1953, hastings1970}, 
transitions of the Markov chain are driven by a random walk and 
exploring the parameter space in this manner can be quite inefficient. Indeed, it is delicate to propose large transitions across the parameter space that will be accepted with high probability. For small transitions, 
converging to the target distribution may require an excessive amount of time. Such methods hence can exhibit low acceptance rates, poor mixing and highly correlated samples \citep{robert2004}. Surged by the development of 
the software package \texttt{Stan} \citep{carpenterstan}, Hamiltonian Monte Carlo \citep[HMC,][]{duane1987, neal2011} has emerged as a reliable alternative for sampling in general settings. The method relies on 
Hamiltonian dynamics to produce large transition paths across the parameter space. 
In practice, the transition paths are approximated using a gradient based 
numerical integrator. In order to preserve the target measure, the numerical solutions must be corrected using an accept-reject step. Both constructing transition paths and the accept-reject step form a 
central issue in this paper as they are unavailable for doubly-intractable distributions. In this paper, we explore the opportunity of using a ``noisy'' version of the HMC sampling method and apply this ``noisy'' 
scheme to the Potts model and to the ERGM.

The paper begins with a review of HMC in Section \ref{sec:background-hmc}. We then introduce the noisy version of the algorithm in Section \ref{sec:noisy-hmc}. The latter relies on Monte Carlo estimates of gradients, 
used in the numerical integration, and unbiased importance sampling estimates of the intractable ratio involved in the accept-reject step. Our approach contrasts with the finite difference schemes or exact derivatives 
usually used in HMC, which are not available. Furthermore, we use all the intermediate points visited by the integrator to derive our importance sampling estimator. This leads to a more robust estimate than the point 
estimates of \cite{murray2006} and extended by \cite{alquier2016} in their work on noisy MCMC. We end the paper with a detailed  numerical study to intractable likelihood problems in Section \ref{sec:numerical-study}.

\section{Background on Hamiltonian Monte Carlo}
\label{sec:background-hmc}

Consider a probability measure $\pi$ on $\bTheta\subset\mathbb{R}^d$ with density, also denoted $\pi$, with respect to the Lebesgue measure, 
\[
\pi(\btheta) = \frac{
\exp\left\lbrace -\pot(\btheta)\right\rbrace
}{
\int_{\bTheta} \exp\left\lbrace 
-\pot(\widetilde{\btheta})
\right\rbrace\dd\widetilde{\btheta}
},
\]
where $\pot$ is continuously differentiable. Markov chain Monte Carlo provides a very general framework to allow estimation of functionals of the form
\[
\int_{\bTheta} g(\btheta)\pi(\mathrm{d}\btheta),
\]
for some function $g$ by generating a Markov chain $\left(\btheta_n\right)_{n\in\mathbb{N}}$ with transition kernel $P$ which leaves $\pi$ invariant. 
Hamiltonian Monte Carlo also belongs to the MCMC toolbox and involves augmenting the target distribution with an
auxiliary variable $\aux\in\mathbb{R}^d$ , usually referred to as a momentum variable, whose density is a $d$-dimensional normal distribution with mean $\zero$ and 
covariance matrix $\mass$, denoted $\N(\cdot\mid\zero,\mass)$ in what follows. HMC thus samples from the augmented distribution
\[
\widetilde{\pi}(\btheta, \aux) = \pi(\btheta)\N(\aux\mid\zero,\mass),
\]
whose marginal chain in $\btheta$ is the distribution of interest. The method originally appeared in statistical physics \citep{duane1987} before being more widely applied for statistical inference. We refer the reader 
to \cite{neal2011} for a comprehensive review. Consider the unnormalised negative joint log-density 
\[
H(\btheta,\aux) = \pot(\btheta) + \frac{1}{2}\aux^{T}\mass^{-1}\aux.
\] 
The method consists of generating proposals for $\btheta$ based on 
the canonical Hamilton's equations which write with respect to a fictitious time $t$
\begin{equation}
\frac{\dd \btheta}{\dd t} = 
\frac{\partial H}{\partial \aux}= \mass^{-1}\aux
\quad\text{and}\quad
\frac{\dd \aux}{\dd t} = -\frac{\partial H}{\partial \btheta} = -\grad_{\btheta}\pot(\btheta),
\label{eqn:hamilton-eqn}
\end{equation}
where $\grad_{\btheta} = \left[\partial/\partial\theta_1, \ldots, \partial/\partial\theta_d\right]^T$ denotes the gradient operator.
The main reason for relying on such a mechanism is to efficiently explore the target density $\pi$ by proposing a new state far from the current state while preserving the measure $\widetilde{\pi}$ \citep{neal2011}.
In particular, the marginal Markov chain on $\bTheta$ is invariant with respect to the target distribution $\pi$. 

In practice, the differential equations \eqref{eqn:hamilton-eqn} cannot be solved analytically, requiring numerical integration to approximate the solution flow.
The most popular numerical integration scheme, if only for its good tradeoff between accuracy and computational cost, is the second order Störmer-Verlet or leapfrog integrator. Consider a time-step $\step$ and the following transformations
\[
g_{1,\step}:(\btheta,\aux)\rightarrow(\btheta+\step\mass^{-1}\aux,\aux)
~\text{ and }~
g_{2,\step}:(\btheta,\aux)\rightarrow(\btheta,\aux-\step/2\grad_{\btheta}\pot(\btheta)).
\]
The leapfrog integrator yields a map $F_\step:\left(\btheta,\aux\right)\rightarrow \left(\btheta',\aux'\right)$ defined by $F_\step= g_{2,\step}\circ g_{1,\step}\circ g_{2,\step}$.
Put another way, the scheme decomposes into the following three-stage procedure:
\begin{align*}
\widetilde{\aux}  = \aux - \frac{\step}{2}\grad_{\btheta}\pot\left(\btheta\right) 
,\quad
\btheta' = \btheta+\step\mass^{-1}\widetilde{\aux}
,\quad
\aux'  = \widetilde{\aux} - \frac{\step}{2}\grad_{\btheta}\pot\left(\btheta'\right).
\end{align*}
The map $F_\step$ approximates the solution at time $\step$ and to get the approximated solution at a time $t$, one iterates it $L=\lfloor\frac{t}{\step}\rfloor$ times, referred to as number of leapfrog steps.

\begin{algorithm2e}[t]
	\caption{Hamiltonian Monte Carlo (single iteration)}
	\label{algo:hmc}
	\KwIn{the current state of the chain $\btheta:=\btheta_0$, a step size $\step$, a number of Leapfrog steps $L$. }
	\medskip
	
	{\bf draw} $\aux:=\aux_0$ from $\N(\zero,\mass)$\;
	
	\For{$\ell\leftarrow 1$ \KwTo $L$}{
		{\bf compute} $\left\lbrace\btheta_\ell, \aux_\ell\right\rbrace  
		= g_{2,\step}\circ g_{1,\step}\circ g_{2,\step}\left(\btheta_{\ell-1}, \aux_{\ell-1}\right)$\;
	}
	
	\medskip
	{\bf set} $(\btheta',\aux') = \left(\btheta_{L}, -\aux_{L}\right)$ with probability
	$
	1 \wedge \exp\left\lbrace H\left(\btheta_{0}, \aux_{0}\right) - H\left(\btheta_{L}, -\aux_{L}\right) \right\rbrace
	$\;
	{\bf set} $(\btheta',\aux') = \left(\btheta_{0}, -\aux_{0}\right)$ otherwise\;
\end{algorithm2e}
Nevertheless, the approximated flow $F_\step$ does not preserve the measure $\widetilde{\pi}(\dd\btheta,\dd\aux)$. To correct the bias introduced, an accept-reject step is used (see Algorithm \ref{algo:hmc}) which following 
a Metropolis-Hastings algorithm, results in a transition from $(\btheta,\aux)$ to $(\btheta',-\aux')$ accepted with probability
\[
\rho(\btheta,\aux,\btheta',\aux') =
1 \wedge \exp\left\lbrace H\left(\btheta, \aux\right) - H\left(\btheta', \aux'\right) \right\rbrace.
\]
The transition kernel of the Metropolis-Hastings update satisfies detailed balance since the deterministic mapping $T\circ F_{\step}$, where $T:(\btheta,\aux)\rightarrow (\btheta,-\aux)$, is an 
involution on $\bTheta\times\bTheta$ \citep{tierney1998}. It is a direct consequence of time reversibility of the approximated flow $F_\step$ -- as each leapfrog step is reversible by 
negating $\step$ -- and its volume preserving property -- as Jacobians of transformations $g_{1,\step}$ and $g_{2,\step}$ have unit determinant. 

\section{HMC for doubly-intractable distribution}
\label{sec:noisy-hmc}

Consider the target distribution $\pi$ being a Bayesian posterior distribution expressed as
\begin{equation}
\pi\left(\btheta\mid \bx\right) \propto f\left(\bx\mid\btheta\right) p(\btheta),
\label{eqn:posterior}
\end{equation}
where $p(\btheta)$ denotes a prior density on the parameter space $\bTheta$ with 
respect to a reference measure (often the Lebesgue measure of the Euclidean space) and $f(\bx\mid\btheta)$ denotes the likelihood of the observed data $\bx\in\Xset$.
Here we are concerned with the situation where the unnormalised posterior distribution, the right-hand-side of (\ref{eqn:posterior}), is intractable. In particular, 
we focus on likelihood models of the form
\begin{equation}
\mathfrak{P} = \left\lbrace 
f\left(\bx\mid\btheta\right) = 
\frac{\exp\left\lbrace A(\btheta,\bx)\right\rbrace}{Z(\btheta)} 
:= 
\frac{q(\btheta,\bx)}{Z(\btheta)} 
\mid \btheta\in\bTheta\subseteq\mathbb{R}^d, A(\cdot,\bx)\in\mathcal{C}^1(\bTheta)
\right\rbrace,
\label{eqn:intractable-like}
\end{equation}
where the parameter dependent likelihood normalising constant, $Z(\btheta)$, is intractable. 
Gibbs random fields represent such a class of intractable likelihood models and are the focus of Section~\ref{sec:numerical-study}.  
This complication results in what is often termed a doubly-intractable
posterior distribution, since the posterior distribution itself is normalised by the evidence (or marginal likelihood) which is typically also intractable. In this 
context, a direct implementation of HMC is not feasible for two reasons:
\begin{enumerate}
	\item The mapping $g_{2,\step}$, and more precisely the gradient $\grad_{\btheta} \pot(\btheta) = -\grad_{\btheta} \log f(\bx\mid\btheta)-\grad_{\btheta}\log p(\btheta)$ is analytically intractable, see 
	Section \ref{sec:grad-estim}, 
	\item The accept-reject step in Algorithm \ref{algo:hmc} is unavailable for doubly-intractable Bayesian analysis as it requires an evaluation of a ratio of intractable
	normalising constants, see Section \ref{sec:MH-estim}.
\end{enumerate}
In what follows, we propose to overcome these two issues by considering Monte Carlo estimates of both the gradient of the log target and the ratio of intractable
normalising constants. Moreover, both of these quantities can be estimated by simulating from the likelihood model, as we will now show in detail. 

\subsection{Gradient estimates}
\label{sec:grad-estim}

Closed-form gradients for complex models are typically out of reach. Computing the gradient in $g_{2,\step}$ is usually addressed using automatic differentiation as in the software package 
\texttt{Stan} \citep{carpenterstan}.  However point-wise estimation is impossible for the likelihood model described in \eqref{eqn:intractable-like} and therefore we require another 
approach. Here we note that the gradient of the log-posterior distribution \eqref{eqn:posterior} can be written as
\begin{equation}
\grad_{\btheta}\log\pi(\btheta\mid\bx) = 
\grad_{\btheta}A(\btheta, \bx) - \grad_{\btheta}\log Z(\btheta) + \grad_{\btheta}\log p(\btheta),
\label{eqn:grad-post}
\end{equation}
Forward-simulations from the likelihood taken at each leapfrog step can be used to provide a Monte Carlo estimate of the gradient, using the following identity,
\begin{align}
\grad_{\btheta}\log Z(\btheta) & = \frac{1}{Z(\btheta)}\grad_{\btheta} Z(\btheta) \nonumber\\
& = \frac{1}{Z(\btheta)}\grad_{\btheta} \int_{\Xset}\exp\left\lbrace A(\btheta,\bx)\right\rbrace\mu(\dd\bx) \nonumber\\
& = \int_{\Xset} \frac{\exp\left\lbrace A(\btheta,\bx)\right\rbrace}{Z(\btheta)} \grad_{\btheta}A(\btheta,\bx)\mu(\dd\bx) \nonumber\\
& = \esp_{\btheta}\left\lbrace \grad_{\btheta} A(\btheta,\bX)\right\rbrace.
\label{eqn:z-as-expect}
\end{align}
So far, we have only assumed that $A(\cdot,\bx)$ is continuously differentiable on $\bTheta$. However this identity holds under regularity conditions which allow one 
to switch the derivative and integral operators (the domain $\Xset$ of $\bX$ is assumed to be independent of $\btheta$) and under the assumption that 
$\grad_{\btheta}A(\btheta,\bX)$ is integrable with respect to $f(\bx\mid\btheta)\mu(\dd\bx)$.
Using Monte Carlo samples $\left\lbrace \bu^{(1, \btheta)},\ldots,\bu^{(N, \btheta)}\right\rbrace$ from $f(\cdot\mid\btheta)$, the expected 
value \eqref{eqn:z-as-expect}
can be estimated using the empirical mean of the random variable $\grad_{\btheta}A(\btheta,\bX)$ over the sample. This leads to the following estimate of the gradient of the log-posterior \eqref{eqn:grad-post} at $\btheta$ 
\begin{equation}
\gradE{N}{\btheta} \log\pi(\btheta\mid\bx) :=
\grad_{\btheta} A(\btheta,\bx) - \frac{1}{N}\sum_{n=1}^{N}\grad_{\btheta}A\left(\btheta,\bu^{(n,\btheta)}\right) + \grad_{\btheta}\log p(\btheta).
\label{eqn:grad-estim}
\end{equation}

\subsection{Metropolis-Hastings ratio estimates}
\label{sec:MH-estim}

The intractability of the likelihood model in (\ref{eqn:intractable-like}) implies, in particular, that the standard MCMC toolbox is infeasible. For example, a naive 
implementation of Algorithm \ref{algo:hmc} when proposing to move from $(\btheta,\aux)$ to $(\btheta',\aux')$ requires the computation of the unknown normalising 
constants, $Z(\btheta)$ and $Z(\btheta')$,
\begin{equation}
\rho\left(
\btheta, \aux, \btheta', \aux'
\right) = 
1 \wedge
\frac{
	Z\left(\btheta\right)
}
{
	Z\left(\btheta'\right)
} 
\frac{
	q\left(\btheta',\bx\right)
}
{
	q\left(\btheta,\bx\right)
}
\frac{
	\N(\aux'\mid\zero,\mass)
}
{
	\N(\aux\mid\zero,\mass) 
}
\frac{
	p(\btheta')
}
{
	p(\btheta)
}.
\label{eqn:alpha-hmc}
\end{equation}

The exchange algorithm \citep{murray2006}, extending the work of \cite{moller2006}, is a popular MCMC method to allow sampling from doubly-intractable 
distributions. 
Denote $\nu(\btheta'\mid\btheta)$ the proposal distribution to move from $\btheta$ to $\btheta'$, the exchange algorithm samples from an augmented distribution
\[
\pi(\btheta', \bu', \btheta \mid \bx) \propto f(\bx\mid\btheta)p(\btheta)\nu(\btheta'\mid\btheta)f(\bu'\mid\btheta').
\] 
whose marginal distribution in $\btheta$ is the posterior distribution of interest. \cite{murray2006} present a clever Metropolis-within-Gibbs algorithm to sample from this augmented 
distribution : given the current value $\btheta$, one iteratively samples $\btheta'$ from its full-conditional $\nu(\cdot\mid\btheta)$, then $\bu'$ from the intractable distribution $f(\cdot\mid\btheta')$ and proposes to deterministically swap $\btheta$ and $\btheta'$ through a Metropolis-Hastings step. It turns out that the ratio of intractable normalising constants drops out of the Metropolis acceptance probability
\[
1\wedge
\frac{
\cancel{\textcolor{red}{Z(\btheta)}}
}{
\cancel{\textcolor{red}{Z(\btheta')}}
}
\frac{
\cancel{\textcolor{red}{Z(\btheta')}} q(\btheta,\bu')
}{
q(\btheta',\bu')\cancel{\textcolor{red}{Z(\btheta)}}
}
\frac{
q(\btheta',\bx)\nu(\btheta\mid\btheta')p(\btheta')
}{
q(\btheta,\bx)\nu(\btheta'\mid\btheta)p(\btheta)
}.
\]
\cite{murray2006} point out that the fraction $q(\btheta,\bu')/q(\btheta',\bu')$ which appears above, can be considered as an single sample importance
estimator of $Z(\btheta)/Z(\btheta')$ since it holds that
\begin{equation}
\esp_{\btheta'}\left\lbrace 
\frac{q(\btheta,\bU')}{q(\btheta',\bU')}\right\rbrace = \frac{Z(\btheta)}{Z(\btheta')},
\label{eqn:z-ratio-is}
\end{equation}
where $\esp_{\btheta'}$ is the expectation with respect to $\bU'\sim f(\cdot\mid\btheta')$. In fact \cite{alquier2016}, consider a generalised exchange algorithm based, 
at each step of the algorithm, on an improved unbiased estimate of $Z(\btheta)/Z(\btheta')$ including multiple auxiliary draws with respect to the proposed 
parameter, namely,
\begin{equation}
\frac{\widehat{Z\left(\btheta\right)}}{Z\left(\btheta'\right)} = 
\frac{1}{N}\sum_{n=1}^N \frac{
q\left(\btheta,\bu^{(n,\btheta')}\right)
}{
q\left(\btheta',\bu^{(n,\btheta')}\right)
},
\label{eqn:ISE-ratio}
\end{equation}
where the auxiliary variables $\left\lbrace\bu^{(1,\btheta')},\ldots,\bu^{(N,\btheta')}\right\rbrace$ are drawn from $f(\cdot\mid\btheta')$. However this so-called noisy exchange 
algorithm no longer leaves the target distribution invariant, nevertheless it is possible to provide convergence guarantees that the resulting Markov chain is close in 
some sense to the target distribution. 
An alternative to previous methods presented and one which we do not explore here but is worth mentioning is Russian Roulette sampling \citep{lyne2015} which can 
be used to get an unbiased estimate of $1/Z(\btheta)$.

Nevertheless, the strategy of using such importance sampling estimates (ISE) in HMC framework is questionable. In particular, the importance sampling weights 
$q(\btheta,\bu)/q(\btheta',\bu)$ can lead to unreliable estimates of the intractable ratio $Z(\btheta)/Z(\btheta')$ for large transitions 
between $\btheta$ and $\btheta' \sim\nu(\cdot\mid\btheta)$, see Figure \ref{fig:noisy-ratio-comp}. In what follows, we develop an alternative importance sampling 
estimator which is compatible with the integrator path.

\subsection{Noisy Hamiltonian Monte Carlo}

To deal with the different difficulties of doubly-intractable Bayesian analysis, we derive a noisy version of HMC, see Algorithm \ref{algo:noisy-hmc}. Consider the set of points 
$\{\btheta = \btheta_{0},\ldots,\btheta' = \btheta_{L}\}$ visited by the symplectic integrator in Algorithm \ref{algo:hmc}. At each leapfrog step $\ell$, we perform $N$ auxiliary 
draws with respect to the current parameter value $\btheta_{\ell}$ to compute surrogates of mapping $g_{2,\step}$ using gradient estimates \eqref{eqn:grad-estim}. Auxiliary draws 
can then be reused to compute at not cost the Metropolis-Hastings proposal \eqref{eqn:alpha-hmc} using importance sampling estimator \eqref{eqn:ISE-ratio}. The transition kernel 
from Algorithm \ref{algo:hmc} is then replaced by an approximated version arising from Algorithm \ref{algo:noisy-hmc} based on stochastic estimators. So far there are no 
theoretical guarantees regarding the effect of the approximated kernel on the limiting distribution and mixing properties. Nonetheless \cite{alquier2016} give some theoretical 
results in the particular case of the Langevin algorithm, that is when $L = 1$. In addition, the work of \cite{chen2014} who establish some results when the gradient of the target 
distribution is estimated using mini-batches of the data may prove useful.

\begin{algorithm2e}[!t]
	\caption{Noisy HMC (single iteration)}
	\label{algo:noisy-hmc}
	\KwIn{the current state of the chain $\btheta = \btheta_{0}$, the observed dataset $\bx$, a step size $\step$, a number of Leapfrog steps $L$, a number of auxiliary draws $N$. }
	
	\medskip
	\setstretch{1.35}
	
	\SetAlgoLined
	{\bf draw} $\aux$ from $\N\left(\zero,\mass\right)$\;
	{\bf draw} auxiliary sample $\left\lbrace\bu^{(1,\btheta_0)},\ldots,\bu^{(N,\btheta_0)} \right\rbrace$ from $f(\cdot\mid\btheta_0)$\;
	
	{\bf compute} 
	$
	\aux_{0} = \aux + \frac{\step}{2}\gradE{N}{{\btheta_0}}
	\log\pi(\btheta_0\mid\bx)
	$\;
	
	\For{$\ell\leftarrow 1$ \KwTo $L$}{
		{\bf compute} $
		\btheta_{\ell} = \btheta_{\ell-1} + \step\mass^{-1} \aux_{\ell-1}$\;
		{\bf draw} auxiliary sample $\left\lbrace\bu^{(1,\btheta_\ell)},\ldots,\bu^{(N,\btheta_\ell)} \right\rbrace$ from $f(\cdot\mid\btheta_{\ell})$\;
		\eIf{$\ell<L$}{
			{\bf compute} $
			\aux_{\ell} = \aux_{\ell-1} + \step\gradE{N}{\btheta_\ell}
			\log\pi(\theta_{\ell}\mid\bx)$\;
		}{
		{\bf compute} $
		\aux_{\ell} = \aux_{\ell-1} + \frac{\step}{2}\gradE{N}{\btheta_\ell}
		\log\pi(\theta_{\ell}\mid\bx)$\;
	}
	\setstretch{1.}
	{\bf compute}
	\[
	\widehat{\frac{Z\left(\btheta_{\ell-1}\right)
	}{
	Z\left(\btheta_{\ell}\right)}} 
	=  \frac{1}{N} \sum_{n=1}^N
	\frac{
	q\left(\btheta_{\ell-1}, \bu^{(n,\btheta_\ell)}\right)
	}{
	q\left(\btheta_{\ell},\bu^{(n,\btheta_\ell)}\right)}
	\]
}
{\bf compute}\vspace{-0.7cm}
\[
\widehat{\rho}\left(\btheta_{0},\aux,\btheta_{L},\aux_{L}\right) = 
1\wedge
\frac{
q(\btheta_{L},\bx)\N\left(\aux_{L}\mid\zero,\mass\right) p\left(\btheta_{L}\right)
}{
q(\btheta_{0},\bx) \N\left(\aux\mid\zero,\mass\right) p\left(\btheta_{0}\right)
}
\prod_{\ell=1}^{L} \frac{
\widehat{Z\left(\btheta_{\ell-1}\right)}
}{
Z\left(\btheta_{\ell}\right)
};
\]
\setstretch{1.35}
{\bf set} $\left(\btheta', \aux'\right) = \left(\btheta_{L},-\aux_{L}\right)$ with probability $\widehat{\rho}\left(\btheta_{0},\aux,\btheta_{L},\aux_{L}\right)$\;
{\bf Otherwise} set $\left(\btheta', \aux'\right) = \left(\btheta_{0},-\aux\right)$\;
\end{algorithm2e}

\begin{figure}[t!]
\centering
	\begin{tikzpicture}[scale=1]
	\node (first){\includegraphics{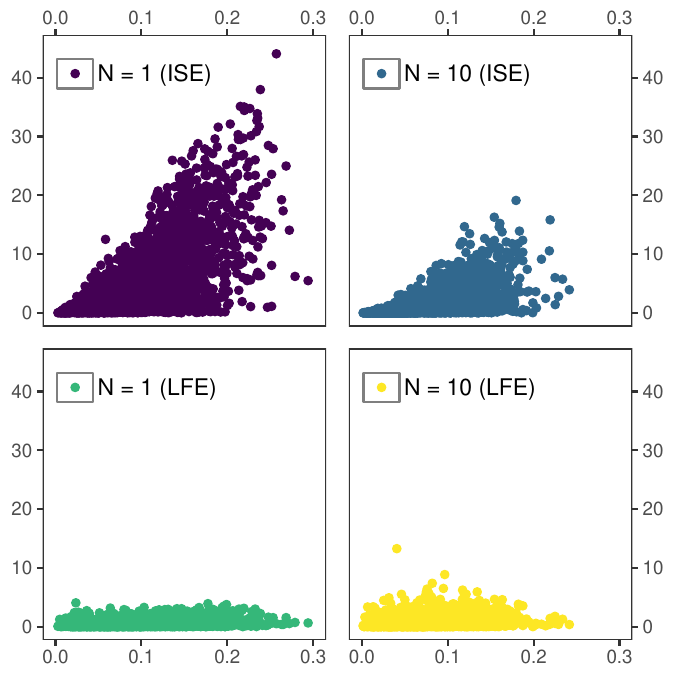}};
	\node[left = of first, rotate=90, xshift = 0.cm, yshift = -0.3cm, anchor = north] {$\mathrm{abs.err}\left(\btheta,\btheta'\right)$};
	\node[below = of first, yshift = 1.3cm]{$\Vert\btheta-\btheta'\Vert$};
	\end{tikzpicture}
	\caption{\small Absolute error of the log-ratio estimates ($y$-axis) with respect to the distance between the current and the proposed states ($x$-axis). The first row refers to 
	importance sampling estimates (ISE) which only use the current and the proposed states. The second row refers to leapfrog estimates (LFE) which make a use of all parameter values 
	involved in the integrator scheme. In terms of absolute error, LFE provides more accurate ratio estimates regardless the distance between the current and the proposed states.}
	\label{fig:noisy-ratio-comp}
\end{figure}

In Algorithm \ref{algo:noisy-hmc}, different strategies can be adopted to estimate the acceptance probability \eqref{eqn:alpha-hmc}.
An obvious solution would be to 
plug in the ISE \eqref{eqn:ISE-ratio} which solely uses auxiliary draws with respect to the proposed value $\btheta' = \btheta^{(L)}$ and discard all others. Nevertheless, such a solution 
turns out to be quite inefficient for large transition. For an illustrative purpose, consider the absolute error of the log ratio estimates
\[
\mathrm{abs.err}\left(\theta,\theta'\right) = \left\vert 
\log\frac{Z(\btheta)}{Z(\btheta')} 
- 
\log\frac{\widehat{Z(\btheta)}}{Z(\btheta')}
\right\vert
\]
at each iteration of HMC. We examined the latter for a particular distribution in $\mathfrak{P}$, namely a Potts model (see Section \ref{sec:mrf} for details) defined on a regular lattice for which 
we can compute exactly the log ratio $Z(\btheta)/Z(\btheta')$ using the R-package \texttt{GiRaF} \citep{giraf}. Figure \ref{fig:noisy-ratio-comp} shows absolute errors obtained for 3,500 pairs 
$(\btheta,\btheta')$ with respect to the distance in $L^2$-norm between the current value $\btheta$ and the proposed one $\btheta'$. Two conclusions can be drawn from first row of Figure \ref{fig:noisy-ratio-comp}. On one hand, a somewhat naive conclusion but consistent with the theory is that the importance sampling estimate is all the more precise that we use multiple auxiliary draws. 
Table \ref{tab:mse-log-ratio} presenting the mean squared error of the estimator emphasises this. On the other hand, the quality of the importance sampling estimate strongly decreases when the 
$L^2$-norm $\Vert\btheta-\btheta'\Vert$ increases. Such a peculiarity has no impact on Metropolis-Hastings methods based on random walks such as the exchange algorithm. Indeed, the variance of the 
proposal is picked to ensure an acceptance rate high enough. This results in proposing move whose norm $\Vert\btheta-\btheta'\Vert$ then remains close to zero where the absolute error is moderate. 
However the deterministic proposals of HMC algorithms are designed to produce large transitions. We thus observe a deterioration of the estimator as shown in Figure \ref{fig:noisy-ratio-comp} 
when $\Vert\btheta-\btheta'\Vert$ increases and therefore the importance sampling estimator \eqref{eqn:ISE-ratio} cannot be advocated in the present situation.
\begin{table}[b!]
	\caption{\small Mean squared error of the log ratio estimator}
	\label{tab:mse-log-ratio}
	\renewcommand{\arraystretch}{1.2}
	\centering
	\begin{tabular}{l|cc}
		& $N = 1$ & $N = 10$ \\
		\hline
		\rowcolor[gray]{0.85} Importance sampling estimator (ISE) & 55.5 & 6.08 \\
		Leapfrog estimator (LFE) & 0.59 & 1.13 \\
	\end{tabular}
\end{table}

To overcome such an issue, we take advantage of all the auxiliary draws with respect to intermediate points $\{\btheta = \btheta_{0},\ldots,\btheta' = \btheta_{L}\}$. The ratio of the normalising 
constant is replaced by an unbiased importance sampling estimate, referred to as the leapfrog estimator (LFE), based on the product of ratios taken at two consecutive points of the integration path, namely
\begin{equation}
\frac{\widehat{Z\left(\btheta\right)}}{Z\left(\btheta'\right)}
=
\prod_{\ell=0}^{L-1} 
\frac{
\widehat{Z\left(\btheta_{\ell}\right)}
}{
Z\left(\btheta_{\ell+1}\right)} 
= \prod_{\ell=0}^{L-1} \frac{1}{N} \sum_{n=1}^N
\frac{
q\left(\btheta_{\ell},\bu^{(n,\btheta_{\ell+1})}\right)
}{
q\left(\btheta_{\ell+1},\bu^{(n,\btheta_{\ell+1})}\right)
},
\label{eqn:LFE-ratio}
\end{equation}
where $\bu^{(\cdot,\btheta_\ell)}$ is sampled from $f(\cdot\mid\btheta_{\ell})$. As it simply reuses the draws involved in gradient estimates \eqref{eqn:grad-estim}, this new estimator comes at no extra 
simulation cost but requires one to evaluate the function $q$ $2N\times L$ times . 
The solution we advocate is slightly different than the exchange algorithm with bridging \citep{murray2006} which relies on an annealed importance sampling technique \citep{neal2001}. Here we do not use 
a sequence of distributions bridging between the original proposal and target distributions for fixed $(\btheta,\btheta')$.

Overall, we observe in Figure \ref{fig:noisy-ratio-comp} that the leapfrog estimators (LFE, second row) are much more accurate than importance sampling estimators (ISE, first row) even 
for a small number of auxiliary draws and thus a poor estimation of the gradient (see Section \ref{sec:grad-estim}). Mean squared errors presented in Table \ref{tab:mse-log-ratio} support this conclusion. 

\subsection{Tuning the noisy HMC algorithm}
\label{sec:tuning}

Tuning HMC often turns out to be a delicate task. 
In practice, sampling from a density $\pi$ using HMC is highly sensitive to user-specified parameters: the step size $\step$, the number of leapfrog steps $L$ and the covariance matrix $\mass$. We refer the 
reader to \cite{neal2011} and \cite{hoffman2014} for a more comprehensive discussion. While the tuning issue is not really the focus of this paper, hereafter we present a 
tuning strategy which we have 
followed for doubly-intractable problems and which could well be improved upon in various ways.

For standard HMC, too large a step size $\step$ results in 
an inaccurate approximated flow $F_{\step}$ and subsequently a high rejection rate. Conversely, if $\step$ is too small, the leapfrog integrator will be precise but will require a significant computational cost 
to simulate a trajectory.
These general considerations should be put in perspective with the quality of the Monte Carlo estimate \eqref{eqn:grad-estim} used in place of the gradient of the log-posterior. Following a poor gradient estimate for too long will affect the dynamic and significantly cut down the acceptance rate. Increasing $N$ will improve the precision in the gradient estimation allowing larger steps but involves increased CPU time sampling from the model. Therefore, a trade-off has also to be found with the number of auxiliary draws $N$. Following the guidelines of \cite{hoffman2014} and also \texttt{Stan} \citep{carpenterstan}, 
the parameter $\step$ was chosen using the dual averaging method, 
that is $\step$ is adaptively tuned during a burn-in period so that for a given integration time $t = \step L$ the average acceptance probability of HMC reaches an optimal value $\delta$. \cite{beskos2013} showed that for a given integration time $t$, the optimal value of $\step$ produces a chain with probability $\delta = 0.65$, approximately. It is not clear whether it is an appropriate probability to target for the noisy HMC algorithm since, for a fixed $N$, the transition kernel of the standard HMC method has been replaced by an approximate version. However since the estimators \eqref{eqn:grad-estim} and \eqref{eqn:LFE-ratio} converge, almost surely, as $N$ goes to infinity, the latter is then the limiting probability associated to the standard kernel and can be used as a rough but reasonable target. 

Furthermore $L$, or equivalently $t$, needs to be large enough
to avoid a random walk behaviour and thereby the slow mixing issue which one would like to prevent in the first place. 
Too long a trajectory is counter-productive since the dynamics retraces its steps bringing the proposed value $\btheta'$ back to a neighbourhood of the current value $\btheta$. 
The NO-U-Turn Sampler \citep[NUTS,][]{hoffman2014} was specifically introduced to avoid such a scenario by producing automatically ``optimal'' trajectory lengths. 
Whilst NUTS fulfils detailed balance and reversibility, this is marred by a major point for doubly-intractable 
distribution. Indeed, NUTS introduces a slice variable whose conditional distribution given $(\btheta, \aux)$ is uniform on $[0;f(\bx\mid\btheta)p(\btheta)\N(\aux\mid\zero,\mass)]$ which requires evaluating the intractable normalising 
constant $Z(\btheta)$. NUTS hence cannot be used to tune $L$ and we have to rely on our personal expertise using a fixed integration time $t$ instead based on the following heuristic. Given a parameter $(\btheta,\aux)$ 
and a step size $\step$, the proposed value $\btheta'$ at the end of Algorithm \ref{algo:hmc} is written as
\[
\btheta' = \btheta + \frac{\step^2L\mass^{-1}}{2}\grad_{\btheta}\log\pi(\btheta\mid\bx)
+\,
\underbrace{\step^2\mass^{-1}\sum_{\ell=1}^{L-1}(L-\ell)\grad_{\btheta}\log\pi\left(\btheta^{(\ell)}\mid\bx\right)}_{:=\Delta(\btheta,\aux)} 
+\, \step L\mass^{-1}\aux.
\]
Given $\btheta$, the latter can be seen as a non-linear transformation of the resampled auxiliary variable $\aux\sim\N(\zero,\mass)$ whose distribution is typically intractable when $L> 1$ due to the term $\Delta(\btheta,\aux)$. We remark that when $L=1$, the $\Delta(\btheta,\aux)$ term vanishes and we get the MALA proposal \citep{roberts1996}. In order to calibrate the integration time, we neglect the randomness of the $\Delta(\btheta,\aux)$ term and approximate the HMC proposal by  
\begin{equation}
\btheta'\mid\btheta\sim\mathcal{N}\left(
\btheta 
+ \frac{\step^2L\mass^{-1}}{2} {\grad_{\btheta}\log\pi(\btheta\mid\bx)},
\step^2L^2\mass^{-1}
\right).
\label{eqn:proposal-hmc}
\end{equation}
The gold standard for dealing with doubly-intractable distributions is the exchange algorithm. For this particular random-walk Metropolis algorithm, the proposal distribution is of the form
\[
\btheta'\mid\btheta \sim \mathcal{N}\left(\btheta,\sigma^2\mass^{-1}\right).
\]
For such a class of algorithms, the optimal scaling \citep{roberts2001} is obtained for 
\begin{equation}
\sigma = \frac{2.38}{\sqrt{d}}.
\label{eqn:optimal-scale}
\end{equation}
In what follows, we set the integration $t$ to the above \eqref{eqn:optimal-scale}. Such a choice ensure that the scaling of the covariance matrix in HMC proposal \eqref{eqn:proposal-hmc} is exactly the same than for the exchange algorithm. This settings is obviously arguable since it leads to a sub-optimal choice for the step-size $\step$. Indeed, the optimal tuning for HMC in high dimension is a step-size scaled as $\step = \ell\times d^{-1/4}$ for some positive constant $\ell$ \citep{beskos2013}. However the current choice leads a step-size scaled as $\step = \ell\times d^{-1/2}$ and can obviously be improved. However being in small dimension, we observed a good behaviour for such a setting.

Finally the last tunable parameter is the mass matrix $\mass$. A mass matrix well suited to the covariance $\Sigma$ of the posterior, namely $\mass =\Sigma^{-1}$, can enhance both the speed and the mixing of HMC. 
For doubly-intractable distributions it is possible to estimate $\Sigma$ in a burn-in phase by using a stochastic approximation algorithm which makes use of the gradient of the log posterior distribution defined in \eqref{eqn:grad-estim}. The method works by estimating the mode $\btheta^{\ast}$ of the posterior \eqref{eqn:posterior} using the gradient \eqref{eqn:grad-estim} within a Robbins Monro algorithm \citep{munro1951} or Ruppert-Polyak averaging \citep{ruppert1988, polyak1992}. Once $\btheta^\ast$ has been found, consequently to Equation \eqref{eqn:z-as-expect} the Hessian of the log posterior at $\btheta^{\ast}$ can be approximated using the sample covariance of auxiliary draws at the mode 
\begin{equation}
\grad_{\btheta}^2 A(\btheta^\ast,\bx) - Cov\left[
\grad_{\btheta} A\left(\btheta^\ast,\bu^{(1,\btheta^\ast)}\right),
\ldots,
\grad_{\btheta} A\left(\btheta^\ast,\bu^{(N,\btheta^\ast)}\right)
\right] + \grad^2_{\btheta}\log p(\btheta^\ast) := -\widehat{\Sigma^{-1}}.
\label{eqn:hes-estim}
\end{equation}
where $\left\lbrace \bu^{(1,\btheta^\ast)},\ldots,\bu^{(N,\btheta^\ast)}\right\rbrace$ are samples from $f(\cdot\mid\btheta^\ast)$. This approximation to the Hessian can hence be used to provide an estimate of $\mass$. We consider this estimation of $\mass$ for Potts and the exponential random graph model numerical study below.

For a matrix $\mass$ that depends on $\btheta$, \cite{girolami2011} provide a fully automated scheme based on the Riemann geometry of the parameter space to adapt $\mass$ along the run. However, $\mass$ then becomes parameter dependent and the leapfrog integrator needs to be replaced by a more sophisticated integration scheme. The latter is somewhat delicate to implement for doubly-intractable distributions and we will not consider it for our numerical study in Section \ref{sec:numerical-study}.

\section{Numerical results}
\label{sec:numerical-study}

\subsection{Toy example: Potts model study}
\label{sec:mrf}

Consider an undirected graph $\G$ inducing a topology on a finite set of sites $\s = \{1,\dots,n\}$: by definition, sites $i$ and $j$ are neighbours (denoted $\ivj$) if and only if $i$ and $j$ are linked by an edge in $\G$. One then calls clique a subset of $\s$ where all elements are mutual neighbours. A discrete Markov random field with respect to undirected graph $\G$ is a random process $\bX = (X_1,\ldots,X_n)$ indexed by $\s$, and taking values in $\Xset \subset \mathbb{Z}^n$, whose conditional distribution of $X_i$, $i\in\s$, depends only upon its neighbours in $\G$.
The Hammersley-Clifford theorem states that if the distribution of a Markov random field with respect to a graph $\G$ is positive for all configurations $\bx$ then it admits a Gibbs representation for the same topology (see \eg \cite{grimmett1973, besag1974} and for a historical perspective \cite{clifford1990}),  
namely a density function $f(\cdot\mid\btheta)$ on $\Xset$ parametrised by $\btheta\in\bTheta\subset\mathbb{R}^d$ and given with respect to the counting measure by
\begin{equation*}
f(\bx\mid\btheta) = \frac{1}{Z\left(\btheta\right)}\exp\left\lbrace A_{\G}(\btheta,\bx)\right\rbrace,
\end{equation*}
where $A_\G$ denotes the potential function which can be written as a sum over the set $\mathcal{C}$ of all cliques of the graph, namely $A_\G(\btheta,\bx) = \sum_{c\in\mathcal{C}}A_c(\btheta,\bx)$ for all configurations $\bx\in\Xset$. The inherent difficulty of all these models arises from the intractable normalising constant, sometimes called the partition function, defined by
\[
Z(\btheta)  = \sum_{\bx\in \Xset} \exp\left\lbrace A_{\G}(\btheta,\bx) \right\rbrace.
\]
The latter is a summation over the numerous possible realisations of the random field $\bX$, which is of combinatorial complexity and cannot be computed directly (except for small grids and states space $\Xset = \{0,\ldots, K-1\}^n$ with small number of states $K$).

In what follows, we focus on a particular pairwise model representative of the general level of difficulty, namely the Potts model \citep{potts1952}
originally used in statistical mechanics to model interacting spins. 
The function $A_\G$ writes as a sum over the cliques of size 1 (corresponding to the nodes of $\G$) and cliques of size 2 (corresponding to the edges of $\G$). Denote $\btheta =(\alpha_0,\ldots,\alpha_{K-1}, \beta)$, 
\[
A_\G(\btheta,\bx) = \sum_{i = 1}^n \sum_{k=0}^{K-1}\alpha_{k}\ind\{x_i = k\} + \beta\sum_{\ivj}\ind\{x_i = x_j\},
\]
where the above sum $\sum_{\ivj}$ ranges the set of edges of the graph $\G$. The parameter $\alpha$ can be interpreted as an external field while the parameter $\beta$ can be interpreted as the inverse of a temperature adjusting the level of dependency between adjacent sites. In the absence of an external field, when the temperature drops below a fixed threshold, called phase transition, the model exhibits strong dependence between neighbors and values $x_i$ of a typical realisation of the field are almost all equal. Note that a potential function on nodes can be defined up to an additive constant. To ensure that potential functions on nodes are uniquely determined, one usually imposes the constraint  $\sum_{k=0}^{K-1}\alpha_k =0$. The dimension of the parameter space $\bTheta$ is then $K$.



\begin{figure}[t]
\centering
\begin{minipage}[t]{7cm}
\centering
\begin{tikzpicture}[scale=0.7]
\foreach \bx in {0,...,4}{
  \draw[dashed, line width=1.5pt, gray!95] (\bx,0) to[out=90,in=-90] (\bx,4);
  \draw[dashed, line width=1.5pt, gray!95] (0,\bx) to[out=0,in=180] (4,\bx);
}

\foreach \bx in {0,...,4} 
	\foreach \y in {0,...,4}
   		\draw[fill = gray!5] (\bx,\y) circle (1.7mm); 

\draw[fill = black] (2,2) circle (1.7mm); 
\draw[fill = gray!70] (2,3) circle (1.7mm); 
\draw[fill = gray!70] (1,2) circle (1.7mm); 
\draw[fill = gray!70] (3,2) circle (1.7mm); 
\draw[fill = gray!70] (2,1) circle (1.7mm); 

\end{tikzpicture}\\
(a)
\end{minipage}
\vspace{0.5cm}
\begin{minipage}[t]{7cm}
\centering
\includegraphics[scale=0.5]{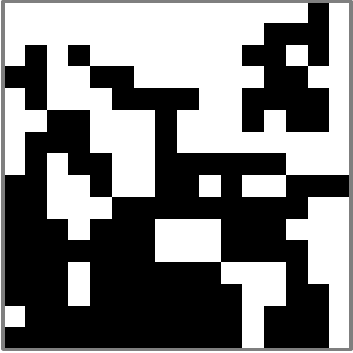}\\
(b)
\medskip
\end{minipage}
\caption{\small (a) First order neighbourhood graphs $\G$. (b) a digital 2-states Potts model defined on a $n = 16\times 16$ regular lattice with a first order neighbourhood system.}
\label{fig:neigh}
\end{figure}
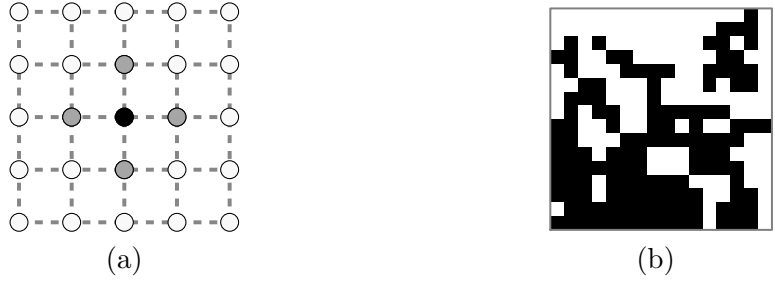

In this example we focus on a digital 2-state Potts model defined on a $16\times 16$ regular lattice with a first order neighbourhood system, see 
Figure \ref{fig:neigh}. This example was chosen since it is one for which we can estimate very accurately the underlying posterior distribution of model parameters and therefore
we can use this as a pedagogical example to compare our noisy HMC algorithm to the corresponding noisy exchange algorithms. We also note that this is not a particularly challenging
example because of the size of the grid and the low number of states. As such, it not one which best exemplifies the performance of the noisy HMC algorithm. 


\paragraph{Ground truth} The R-package \texttt{GiRaF} \citep{giraf} allows one to compute exactly the normalising constant $Z(\btheta)$ of a Potts model defined on a rectangular $h\times w$ lattice \citep{friel2007}. 
The algorithm to evaluate $Z(\btheta)$ is exponential in the number of rows $h$ and linear in the number of columns $w$. As such it can handle models defined on a lattice up 
to $h=25$ for $K=2$. For such a lattice, we can then compute 
ground truth quantities against which we can compare the output from the various algorithms that we consider in this paper. In particular using adaptive cubature algorithms from the R-package \texttt{cubature} \citep{cubature} we can compute the posterior mean
\[
\bar{\btheta} = \esp_{\pi}\left\lbrace\btheta\right\rbrace = \left[\int_{\bTheta} \theta_i{\pi(\btheta\mid\bx)}\dd\btheta\right]_i,
\]
and the Kullback-Leibler divergence between the empirical distribution $Q$ and the stationary distribution $\pi$, namely
\[
\mathrm{KL}(Q\Vert\pi) = \int_{\bTheta} Q(\btheta)\log\frac{Q(\btheta)}{\pi(\btheta\mid\bx)}\dd\btheta.
\]
To estimate the latter, we divided the parameter space $\bTheta$ into a set of bins of size $0.01 \times 0.01$. Then, we solely computed integrals over the set of non-empty bins, integrals being set to zero otherwise.

\paragraph{MAP and Hessian estimates} The maximum \textit{a posteriori} (MAP) was estimated using the Ruppert-Polyak averaging method. In this scheme, at each iteration the gradient was estimated using the 
identity \eqref{eqn:grad-estim} with $N = 10$ draws from the likelihood. The algorithm stops when $\Vert\btheta_{n}-\btheta_{n+1}\Vert$ drops below a $1e^{-3}$ threshold. Once the MAP $\btheta^{\ast}$ is estimated, we estimate the 
Hessian $\grad^{2}_{\btheta}\log\pi(\btheta^{\ast}\mid\bx)$ using identity \eqref{eqn:hes-estim} with $N=500$ draws from the likelihood.

\begin{figure}[t!]
	\begin{tikzpicture}[scale=1]
	\node (first){\includegraphics{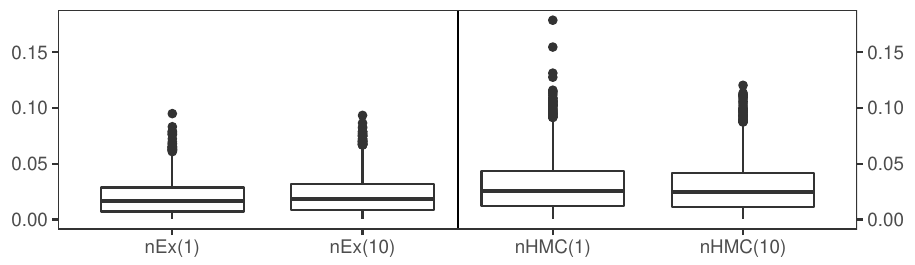}};
	\node[left = of first, rotate=90, xshift = 0.cm, yshift = -0.3cm, anchor = north] {$\vert\alpha-\alpha'\vert$};
	\end{tikzpicture}
		\begin{tikzpicture}[scale=1]
	\node (first){\includegraphics{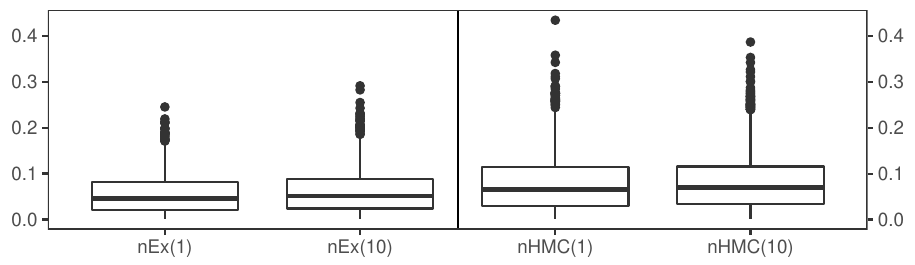}};
	\node[left = of first, rotate=90, xshift = 0.cm, yshift = -0.3cm, anchor = north] {$\vert\beta-\beta'\vert$};
	\end{tikzpicture}
	\caption{\small Potts model: boxplot of absolute accepted moves lengths for each component of $\btheta=(\alpha,\beta)$ in the exchange (nEx(1)), the noisy exchange (nEx(10)) and the noisy HMC (nHMC(1), nHMC(10)). 
	The accepted moves of HMC algorithms are greater than the two random walk exchange algorithms across both parameter directions.}
	\label{fig:move-size}
\end{figure}
\paragraph{First experiment: nHMC(N)} We ran the noisy HMC algorithm for various numbers of draws, namely $N=1$ and $N=10$, in order to compute the gradient estimate \eqref{eqn:grad-estim} and the leapfrog estimator \eqref{eqn:LFE-ratio}. The covariance matrix of the auxiliary variable distribution is set to $\mass = \widehat{\Sigma^{-1}}$. For each value of $N$, the step size $\step$ and the number of leapfrog steps $L$ were tuned as described in Section \ref{sec:tuning} with $\delta = 0.65$. The burn-in chain was of 500 iterations to tune the parameters for each setting. 

 
\paragraph{Second experiment: nEx(N)} The gold standard for conducting doubly-intractable Bayesian inference is the exchange algorithm \citep{murray2006}. We ran the noisy exchange algorithm \citep{alquier2016} for 
various number of auxiliary draws, namely $N=1$ and $N=10$. The proposal distribution is set to be a 2-dimensional normal distribution with mean $\btheta$ and covariance matrix $2.38^{2}d^{-1}\widehat{\Sigma}$ in order 
to target the optimal acceptance probability for Metropolis-Hastings algorithm, namely $\delta = 0.234$ \citep{roberts2001}, as already mentioned in Section \ref{sec:tuning}.

For the various experiments aforementioned, we considered a pseudo-observation exactly drawn from a Potts model using the R-package \texttt{GiRaF} with parameter $\btheta=(0.,0.5)$ and a uniform prior on $\bTheta = [-0.5;0.5]\times[0;1]$. 
Once the parameters were tuned, we ran 20 chains of size 4,500 per experiment. Figure \ref{fig:move-size} shows the typical behaviour of the accepted moves length for both the noisy exchange, nEx($N$), 
and noisy HMC, nHMC($N$) algorithms, when $N=1$ or $N=10$ auxiliary draws. First, we can observe that the noisy version of those algorithms still benefits from using a well chosen mass matrix $\mass$. Indeed, the 
chain is then able to move more in the less constrained direction, namely $\beta$. On the other hand, both algorithms produce a proposal with the same covariance terms but the HMC chain manages to move more across the 
parameter space despite a sub-optimal choice of $\step$. 
We surmise that the performance could be further improved with a better tuning for $\step$.

\begin{table}[b!]
\caption{\small Potts model: noisy HMC, exchange and noisy exchange algorithms outputs average (standard deviation) over 20 chains of size 4,500 for various number of auxiliary draws ($N=1$ and $N=10$).}
\label{tab:ap}
\renewcommand{\arraystretch}{1.2}
\centering
\rowcolors{1}{}{gris85}
\begin{tabular}{l|cccc}
\hline
 & nHMC(1) & nHMC(10)  & nEx(1) & nEx(10) \\
 \hline
 \hline
$\rho\left(\theta,\theta'\right)$ (\%) & 62.8 (1e-2) & 63.5 (8.2e-3)  & 22.5 (7.5e-3) & 28.9 (7.7e-3) \\
$\epsilon$ & 0.13 & 0.80  & \multicolumn{2}{c}{NA}\\ 
L & 12 & 2  & \multicolumn{2}{c}{NA}\\ 
Running time (s) & 1456 (110) & 3268 (34.4)  & 110.7 (4.4) & 1133 (34) \\
ESS($\alpha$) & 1604 (103) & 1891 (169)  & 244.4 (34.7) & 394.8 (40.8) \\
ESS($\alpha$) per sec.& 1.12 (9.1e-2) & 0.57 (5.3e-2)  & 2.21 (0.34)  & 0.35 (3.8e-2) \\
ESS($\beta$)  & 1573 (181) & 1904 (186)  & 241.4 (23.6) & 396.1 (41.4)  \\
ESS($\beta$) per sec.& 1.09 (0.13) & 0.59 (5.7e-2)  & 2.18 (0.24) & 0.35 (3.8e-2) \\
MSE($\bar{\btheta})$ ($\times$1e-5) & 0.34 & 1.68  & 3.80 & 1.52 \\
KL-divergence & 0.23 (1.8e-2)  & 0.18 (1.3e-2)   & 0.43 (5.1e-2)  & 0.33 (2.5e-2)\\
\end{tabular}
\end{table}

Table \ref{tab:ap} summarises different outputs average obtained over these chains. As a first step, we note that the dual averaging scheme with a Monte Carlo estimate \eqref{eqn:grad-estim} of the gradient 
still ensures that the average Metropolis acceptance probability $\rho(\btheta,\btheta')$ is close to its required value. As regards the value $\step$ and $L$, we observe a trade-off with the quality of the 
gradient as already mention in Section \ref{sec:tuning}.  In order to keep proposing transitions which are accepted with high probability, a poor gradient estimate leads to numerous small steps, therefore 
avoiding to follow an incorrect gradient for too long, and conversely, $\step$ becomes larger and $L$ smaller when we put more computation effort to get a more accurate gradient estimate. 

The major motivation for using HMC was to have better mixing. Table \ref{tab:ap} gives Effective Sample Size (ESS) with respect to the marginal chain in $\alpha$ (denoted $\mathrm{ESS}(\alpha)$) and the 
marginal chain in $\beta$ (denoted $\mathrm{ESS}(\beta)$) so as to stress that difference of behaviour between noisy algorithms. In terms of a per iteration basis, noisy HMC overall mixes better than for 
various exchange algorithms, as one would expect. However it is significantly more expensive in terms of computation time. Indeed, the cost of the different algorithms is largely determined by the number of 
auxiliary draws from the model. HMC performs  $N\times(L+1)$ draws per iteration whereas the noisy exchange does $N$ draws. On a per unit of computational time basis, the exchange algorithm is then typically 
superior to the noisy HMC algorithms, at least in terms of ESS.
We further remark here that the noisy version proposed by \cite{alquier2016} comes at a loss in practice as the overall ESS does not increase significantly enough against the extra computational cost. 

On a per iteration basis, noisy HMC present better results in terms of the mean squared error of the posterior mean, namely $\mathrm{MSE}(\bar{\btheta})$, and Kullback-Leibler divergence. One could argue that 
such results might not hold on a per unit of computational time basis as we would have a chain 10 times longer for the exchange algorithm and therefore expect to have better result in terms of 
$\mathrm{MSE}(\bar{\btheta})$ and Kullback-Leibler divergence, though it is not completely obvious whether this would be the case.

\subsection{Exponential random graph model study}
\label{sec:ergm}

\begin{figure}[b!]
	\centering
	\includegraphics[width=0.5\textwidth]{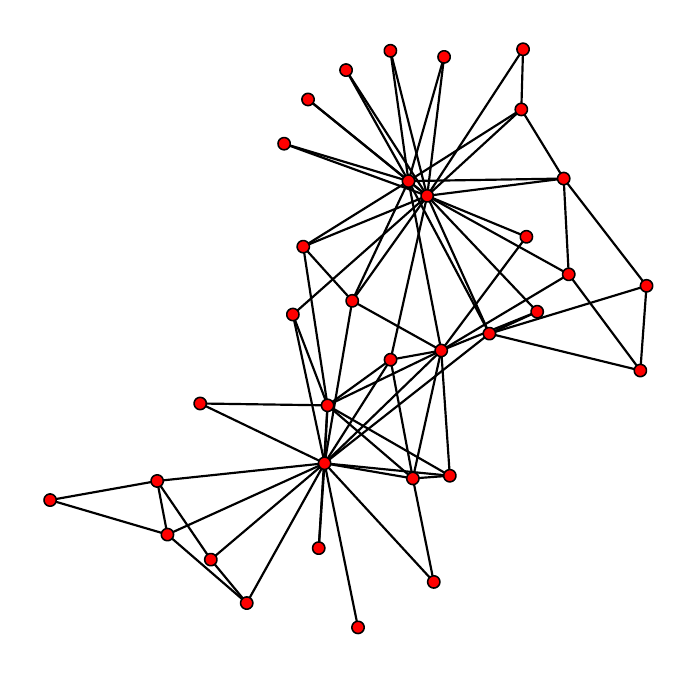}
	\caption{\small Zachary karate club: A social network of 34 individuals in a karate club. The network has 78 edges and 528 2-stars.}
	\label{fig:karate}
\end{figure}

Exponential random graph models are a family of network models that generalise Markov random graphs.
A graph is a collection of $n$ nodes connected by edges. The edges are indicated by a $n \times n$ adjacency matrix $\bx$ where
\begin{equation*}
x_{ij} =
\begin{cases}
1, \hspace{1em}\text{if node $i$ and node $j$ are connected},\\
0, \hspace{1em}\text{otherwise}.
\end{cases}.
\end{equation*}
These edges may be directed or undirected, in the latter case $x_{ij} = x_{ji}$ for all pairs $(i,j)$. An example of such a network is shown in Figure~\ref{fig:karate} depicting the friendship connections of 34 individuals (nodes) in a karate club \citep{karateclub}.

The energy function $A(\btheta,\bx)$ for an exponential random graph model on a directed or an undirected graph $\bx$ with $n$ nodes is,
\begin{equation}
A\left(\btheta, \bx\right) = \sum_{i=1}^{d} \theta_i s_i(\bx),
\label{eqn:ergm_pot}
\end{equation}
where the terms $s_i(\bx)$ are typically sufficient statistics. These statistics capture local structure of the network $\bx$, for example, the count of edges or triangles in the network. The normalising constant for the exponential random graph models involves a sum over all $2^{\binom{n}{2}}$ realisable graphs $\bx \in \Xset$ in the undirected case and $2^{n(n-1)}$ in the directed case,
\[
Z(\btheta)  = \sum_{\bx\in \Xset} \exp\left\lbrace \sum_{i=1}^{d} \theta_i s_i(\bx) \right\rbrace
\]
and is intractable for all but small graphs. 

In this numerical study we consider a two parameter ERGM where the sufficient statistics are $s_1(\bx) = \sum_{i < j} x_{ij}$, counting the number of undirected edges in $\bx$ and 
$s_2(\bx) = \sum_{i} \sum_{j < k} x_{ij}x_{ik}$, counting the number of so-called 2-stars in the network. The sufficient statistics are assigned parameters $\btheta = (\theta_1, \theta_2)$ with model energy function,
\begin{equation}
A\left(\btheta,\bx\right) = \theta_1 s_1(\bx) + \theta_2 s_2(\bx).
\label{eq:zacclub}
\end{equation}

The posterior distribution for the parameters of an exponential random graph model as defined with potential \eqref{eqn:ergm_pot} is doubly-intractable, therefore it is not possible to obtain a a ground truth for the posterior distribution of the parameters in order to compare with HMC. This is in contrast to the Potts model study in which a ground truth is available for small lattices using the R-package \texttt{GiRaF}. Instead using the R-package \texttt{Bergm} \citep{bergm2014}, an exchange algorithm was run for a long period of time in order to sample from the posterior \eqref{eqn:posterior}. This long run consisted of 100{,}000 iterations for the parameter $\btheta$ with a per iteration auxiliary burn in of 1{,}000{,}000 to obtain the estimate the single importance sampling estimate \eqref{eqn:z-ratio-is}.

\paragraph{First experiment: nHMC(N)}
We ran the noisy HMC algorithm (Algorithm \ref{algo:noisy-hmc}) for $N=10$ draws in order to compute the gradient estimate \eqref{eqn:grad-estim} and the leapfrog estimator \eqref{eqn:LFE-ratio}. To tune the algorithm 
the $\mass$ matrix was set equal to an estimate of the negative Hessian of the log-posterior \eqref{eqn:hes-estim}. This approximation was computed to locating the posterior mode $\btheta^\ast$ by iterating the following 
Robbins-Munro scheme for $i = 1, \dots, 200,$
\begin{align*}
\btheta^{(i+1)} = \btheta^{(i)} +\frac{\alpha}{i}\hat{\grad}_{\bu^{(\btheta,N=10)}}\log\pi(\btheta^{(i)}\mid\bx),
\end{align*}
with $\alpha = 1$. The approximate location of the mode is taken as $\btheta^\ast = \btheta^{(201)}$. Then 500 draws are sampled from $f(\cdot\mid\btheta^\ast)$ to compute the negative Hessian estimate $\widehat{\Sigma^{-1}}$ using \eqref{eqn:hes-estim}. The mass matrix $\mass$ was then set equal to $\widehat{\Sigma^{-1}}$. 

\paragraph{Second experiment: nEx(N)}
The exchange algorithm \citep{murray2006} was run for 5{,}000 iterations of the parameter $\btheta$. The noisy exchange \citep{alquier2016} was run for 5{,}000 iterations of the parameter $\btheta$ using $N = 25$ auxiliary draws at each parameter value. The proposal distribution for both the exchange and noisy was taken as a normal distribution with mean at the current $\btheta$ and covariance matrix $2.38^2d^{-1}\mass^{-1}$, where $\mass = \widehat{\Sigma^{-1}}$ is approximated using \eqref{eqn:hes-estim} as with the noisy HMC study.

\begin{table}[b!]
	\caption{\small Zachary karate club: HMC, exchange and noisy exchange algorithms results for the Zachary karate club network. The HMC was performed with 10 draws from the likelihood to estimate the gradients. Each chain is of size 5,000 and replicated 20 times to produce the results in this table.}
	\label{tab:ergm_results}
	\renewcommand{\arraystretch}{1.2}
	\centering
	\rowcolors{1}{}{gris85}
	\begin{tabular}{l|rrr}
		\hline
		 & nHMC(10) & nEx(1) & nEx(25) \\
		\hline
		\hline
		$\rho\left(\theta,\theta'\right)$ (\%) & 61.7 (7e-3) & 23.46 (6e-3) & 29.44 (8e-3)  \\
		$\epsilon$ & $3.453 \times 10^{-3}$ & NA & NA \\  
		L   & 487  & NA & NA\\ 
		Running time (s) & 1622 (92)		 	&  706 (111)	& 	821 (149)	\\
		ESS ($\theta_\text{edge}$) & 3132 (250)	 	&  260 (43)		& 296 (40)		\\
		ESS ($\theta_\text{edge}$) per sec. & 1.94 (0.22)	&  		0.377 (0.08)	& 	0.368 (0.06)	\\
		ESS ($\theta_\text{2star}$) 	& 3379 (267)	 	&  		285 (57)	& 	323 (39)		\\
		ESS ($\theta_\text{2star}$) per sec. & 2.09 (0.23)	&  		0.412 (0.10)	& 	0.404 (0.07)		\\
		MSE (ground truth) &  $1.968 \times 10^{-5}$ & 0.0249 & 0.00974
	\end{tabular}
\end{table}

\begin{figure}[htbp]
	\centering
	\includegraphics[width=15cm]{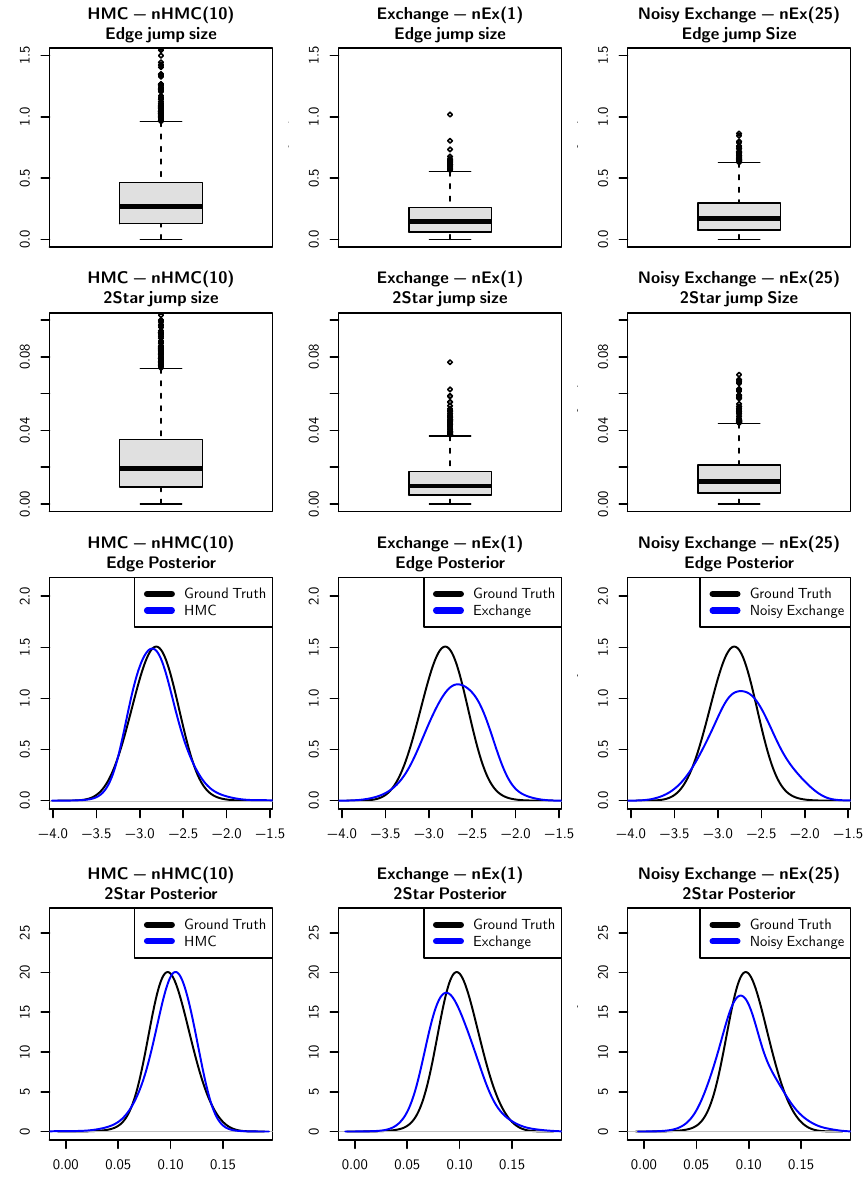}
	\caption{\small Zachary karate club: HMC, exchange and noisy exchange algorithm results. The jump sizes or absolute accepted moves lengths of the HMC algorithm are greater than the two random walk 
	exchange algorithms across both parameter directions. The HMC algorithm shows good convergence to the ground truth results.}
	\label{fig:karate_results}
\end{figure}

The different algorithms were replicated 20 times using a different starting point and random seed for each experiment. The results of the various algorithms run on the Zachary karate club for the model defined by \eqref{eq:zacclub} are shown in Table \ref{tab:ergm_results} and Figure \ref{fig:karate_results}. The noisy HMC provides a noticeable increase in ESS per unit time with respect to the exchange and noisy exchange algorithms. The mean squared error (MSE) to the ground truth value of $\theta$ is also significantly lower for HMC and this is visible in the marginal density estimates given in Figure \ref{fig:karate_results}, where the HMC algorithm matches closely to each marginal posterior distribution from the ground truth.

\section{Conclusions}

The paper yields a noisy version of the HMC algorithm for doubly-intractable Bayesian analysis. This noisy version of the HMC algorithm is based on Monte Carlo estimates for the gradient based leapfrog numerical 
integrator and a novel estimator for the Metropolis-Hastings proposal compatible with the integration scheme. Throughout the numerical studies described here, we have shown that it improves the statistical 
efficiency of the gold standard exchange algorithm which is only capable of exploring the parameter space slowly by making local moves.
The computational cost of sampling from the model is the main bottleneck of computing gradients and importance sampling estimators required in HMC. Further research is required to assess the trade-off between 
the gradient quality, in terms of number of samples from the likelihood, and the efficiency of HMC. 
Noisy HMC opens vistas for further research. 

\paragraph*{Theoretical support} On the first hand, the paper motivates an in-depth theoretical study of noisy HMC. Indeed, the transition kernel of the original HMC method has been replaced by an approximate 
kernel due to Monte Carlo estimates being used within each iteration. This breaks the ergodicity and the noisy HMC may not admit the target posterior distribution $\pi$ as the invariant distribution. 
\cite{alquier2016} have examined the ergodicity for noisy MCMC algorithms and provide bounds on the total variation distance between a Markov chain with the desired target distribution, and the Markov chain of 
a noisy MCMC algorithm. Their noisy MALA is a special case of our noisy HMC when the number of leapfrog steps is $L = 1$. While their results do not hold for larger $L$, one might expect similar conclusion for 
noisy HMC algorithms. 
Nevertheless the intractable HMC proposal forms to some extent a 
hurdle for the proof regarding noisy HMC. Another question that has also been opened by \cite{alquier2016} is the asymptotic variance of estimators from noisy HMC algorithm.

\paragraph*{Optimal tuning} The tuning strategy proposed in this paper is not optimal though the results for the noisy HMC algorithm we presented are already convincing. The overall performances could be improved 
by a better choice of the integration time and in general by introducing an automatic problem-specific strategy to pick $\step$ and $L$. A path to explore is to adapt the NUTS sampler in the context of doubly-intractable Bayesian 
inference. A follow-up to the theoretical support of noisy HMC would also be to provide an optimal setting.

\paragraph*{Latent process} Another specific instance of likelihood intractability occurs when the model relies on a latent process $\bx$ whose state space $\Xset$ is of high dimension. Consider a likelihood expressed 
as a multidimensional integral,
\[
\ell(\btheta \mid\y) = \int_{\Xset} \ell(\btheta\mid\y,\bx)\mu(\mathrm{d}\bx),
\]
and impossible to evaluate. Such problems arise frequently in applied statistics for example hidden Markov models. A fruitful avenue to explore is the development of noisy HMC methods for this class of statistical model. 
In the paper, we propose Monte Carlo estimates based on forward-simulations as surrogates for the gradient based integrator. Likewise, the score function of $\ell(\btheta \mid\y)$ can be estimated using following identity,
\[
\grad_{\btheta}\log\ell(\btheta \mid\y)  = \esp_{\btheta,\y}\left\lbrace \grad_{\btheta}\log\ell(\btheta,\bX\mid\y)\right\rbrace,
\]
where the expectation is with respect to $\ell(\bx\mid\btheta,\y)$ as described in \citet{friel2016}. Moreover, for hidden Markov models, the Metropolis-Hastings acceptance probability can still be evaluated 
using the leapfrog estimator we introduced in this work.

\bibliographystyle{abbrvnat}
\bibliography{biblio}

@article {augustin1998,
author = {Augustin, N. H. and Mugglestone, M. A. and Buckland, S. T.},
title = {The role of simulation in modelling spatially correlated data},
journal = {Environmetrics},
volume = {9},
number = {2},
publisher = {John Wiley & Sons, Ltd.},
pages = {175--196},
year = {1998}
}

@Article{alquier2016,
Title = {Noisy {M}onte {C}arlo: convergence of {M}arkov chains with approximate transition kernels},
Author = {Alquier, Pierre and Friel, Nial and Everitt, Richard and Boland, Aidan},
Journal = {Statistics and Computing},
Year = {2016},
Number = {1},
Pages = {29--47},
Volume = {26}
}

@Article{andrieu2009,
Title = {The {P}seudo-{M}arginal {A}pproach for {E}fficient {M}onte {C}arlo {C}omputations},
Author = {Andrieu, Christophe and Roberts, Gareth O.},
Journal = {The Annals of Statistics},
Year = {2009},
Number = {2},
Pages = {697--725},
Volume = {37}
}

@Article{beaumont2003,
Title = {Estimation of {P}opulation {G}rowth or {D}ecline in {G}enetically {M}onitored {P}opulations},
Author = {Beaumont, Mark A.},
Journal = {Genetics},
Year = {2003},
Number = {3},
Pages = {1139--1160},
Volume = {164}
}

@Article{besag1974,
Title = {Spatial {I}nteraction and the {S}tatistical {A}nalysis of {L}attice {S}ystems (with {D}iscussion)},
Author = {Besag, Julian E.},
Journal = {Journal of the Royal Statistical Society. Series B (Methodological)},
Year = {1974},
Number = {2},
Pages = {192--236},
Volume = {36},
Publisher = {Wiley for the Royal Statistical Society}
}

@Article{beskos2013,
Title = {Optimal tuning of the hybrid {M}onte {C}arlo algorithm},
Author = {Beskos, Alexandros and Pillai, Natesh and Roberts, Gareth and Sanz-Serna, Jesus-Maria and Stuart, Andrew},
Journal = {Bernoulli},
Year = {2013},
Number = {5A},
Pages = {1501--1534},
Volume = {19}
}

@Article{clifford1990,
Title = {Markov random fields in statistics},
Author = {Clifford, Peter},
Journal = {Disorder in physical systems: A volume in honour of John M. Hammersley},
Year = {1990},
Pages = {19--32},
Publisher = {Clarendon Press}
}

@Article{duane1987,
Title = {Hybrid Monte Carlo},
Author = {Duane, Simon and Kennedy, A.D. and Pendleton, Brian J. and Roweth, Duncan},
Journal = {Physics Letters B},
Year = {1987},
Number = {2},
Pages = {216 - 222},
Volume = {195}
}

@article{francois2006,
author = {Fran\c{c}ois, Olivier and Ancelet, Sophie and Guillot, Gilles}, 
title = {Bayesian {C}lustering {U}sing {H}idden {M}arkov {R}andom {F}ields in {S}patial {P}opulation {G}enetics},
volume = {174}, 
number = {2}, 
pages = {805--816}, 
year = {2006}, 
journal = {Genetics} 
}

@InProceedings{friel2012,
Title = {Bayesian {I}nference for {G}ibbs {R}andom {F}ields {U}sing {C}omposite {L}ikelihoods},
Author = {Friel, Nial},
Booktitle = {Proceedings of the Winter Simulation Conference},
Year = {2012},
Number = {28},
Pages = {1--8},
Publisher = {Winter Simulation Conference},
Series = {WSC '12},
publisher = {Winter Simulation Conference}
}

@Article{friel2007,
Title = {Recursive computing and simulation-free inference for general factorizable models},
Author = {Friel, Nial and Rue, H\r{a}vard},
Journal = {Biometrika},
Year = {2007},
Number = {3},
Pages = {661--672},
Volume = {94}
}

@article{friel2016,
author = {Friel, Nial and Mira, Antonietta and Oates, Chris J.},
journal = {Bayesian Analysis},
number = {1},
pages = {215--245},
publisher = {International Society for Bayesian Analysis},
title = {Exploiting {M}ulti-{C}ore {A}rchitectures for {R}educed-{V}ariance {E}stimation with {I}ntractable {L}ikelihoods},
volume = {11},
year = {2016}
}

@Article{girolami2011,
Title = {Riemann manifold {L}angevin and {H}amiltonian {M}onte {C}arlo methods},
Author = {Girolami, Mark and Calderhead, Ben},
Journal = {Journal of the Royal Statistical Society: Series B (Statistical Methodology)},
Year  = {2011},
Number = {2},
Pages = {123--214},
Volume = {73}
}

@Article{green2002,
Title = {Hidden {M}arkov {M}odels and {D}isease {M}apping},
Author = {Green, Peter J. and Richardson, Sylvia},
Journal = {Journal of the {A}merican {S}tatistical {A}ssociation},
Year = {2002},
Number = {460},
Pages = {1055--1070},
Volume = {97}
}

@Article{grimmett1973,
Title = {A theorem about random fields},
Author = {Grimmett, Geoffrey R},
Journal = {Bulletin of the London Mathematical Society},
Year = {1973},
Number = {1},
Pages = {81--84},
Volume = {5},
Publisher = {Oxford University Press}
}

@Article{hastings1970,
Title = {Monte {C}arlo sampling methods using {M}arkov chains and their applications},
Author = {Hastings, W Keith},
Journal = {Biometrika},
Year = {1970},
Number  = {1},
Pages = {97--109},
Volume = {57}
}

@Article{hoffman2014,
Title = {The {N}o-{U}-turn sampler: adaptively setting path lengths in {H}amiltonian {M}onte {C}arlo.},
Author = {Hoffman, Matthew D and Gelman, Andrew},
Journal = {Journal of Machine Learning Research},
Year = {2014},
Number = {1},
Pages = {1593--1623},
Volume = {15}
}

@InCollection{hurn2003,
Title = {A {T}utorial on {I}mage {A}nalysis},
Author = {Hurn, Merrilee~A. and Husby, Oddvar~K. and Rue, H\r{a}vard},
Booktitle = {Spatial Statistics and Computational Methods},
Publisher = {Springer New York},
Year = {2003},
Pages = {87--141},
Series = {Lecture Notes in Statistics},
Volume = {173},
Editor = {Jesper Møller}
}

@Article{jaakkola2000,
Title = {Bayesian parameter estimation via variational methods},
Author = {Jaakkola, Tommi S. and Jordan, Michael I.},
Journal = {Statistics and Computing},
Year = {2000},
Number = {1},
Pages = {25--37},
Volume = {10}
}

@Article{jordan1999,
Title = {An {I}ntroduction to {V}ariational {M}ethods for {G}raphical {M}odels},
Author = {Jordan, Michael I and Ghahramani, Zoubin and Jaakkola, Tommi S and Saul, Lawrence K},
Journal = {Machine learning},
Year = {1999},
Number = {2},
Pages = {183--233},
Volume = {37}
}

@Article{lindsay1988,
Title = {Composite likelihood methods},
Author = {Lindsay, Bruce G},
Journal = {Contemporary Mathematics},
Year = {1988},
Number = {1},
Pages = {221--39},
Volume = {80}
}

@Article{lyne2015,
Title = {On {R}ussian {R}oulette {E}stimates for {B}ayesian {I}nference with {D}oubly-{I}ntractable {L}ikelihoods},
Author = {Lyne, Anne-Marie and Girolami, Mark and Atchadé, Yves and Strathmann, Heiko and Simpson, Daniel},
Journal = {Statistical Science},
Year = {2015},
Number = {4},
Pages = {443--467},
Volume = {30}
}

@Article{moller2006,
Title = {An efficient {M}arkov chain {M}onte {C}arlo method for distributions with intractable normalising constants},
Author = {M{\o}ller, Jesper and Pettitt, Anthony N and Reeves, R and Berthelsen, Kasper K},
Journal = {Biometrika},
Year = {2006},
Number = {2},
Pages = {451--458},
Volume = {93}
}

@Article{marin2012,
Title = {{A}pproximate {B}ayesian {C}omputational methods},
Author = {Marin, Jean-Michel and Pudlo, Pierre and Robert, Christian P. and Ryder, Robin J.},
Journal = {Statistics and Computing},
Year = {2012},
Number = {6},
Pages = {1167--1180},
Volume = {22}
}

@Article{metropolis1953,
Title = {Equation of state calculations by fast computing machines},
Author = {Metropolis, Nicholas and Rosenbluth, Arianna W and Rosenbluth, Marshall N and Teller, Augusta H and Teller, Edward},
Journal = {The journal of Chemical Physics},
Year = {1953},
Number = {6},
Pages = {1087--1092},
Volume = {21}
}

@InCollection{murray2006,
Title = {{MCMC} for doubly-intractable distributions},
Author = {Iain Murray and Zoubin Ghahramani and David J. C. MacKay},
Booktitle = {Proceedings of the 22nd Annual Conference on Uncertainty in Artificial Intelligence (UAI-06)},
Publisher = {AUAI Press},
Year = {2006},
Pages = {359--366}
}

@InBook{neal2011,
Title = {Handbook of Markov Chain Monte Carlo},
Author = {Neal, Radford M},
Chapter = {5},
Pages = {113--162},
Publisher = {CRC press},
Year = {2011},
Opteditor = {Brooks, Steve and Gelman, Andrew and Jones, Galin and Meng, Xiao-Li}
}

@Article{neal2001,
Title = {Annealed importance sampling},
Author = {Neal, Radford M},
Journal = {Statistics and Computing},
Year = {2001},
Number = {2},
Pages = {125--139},
Volume = {11}
}

@InProceedings{potts1952,
Title = {Some generalized order-disorder transformations},
Author = {Potts, Renfrey Burnard},
Booktitle = {Mathematical proceedings of the cambridge philosophical society},
Year = {1952},
Number = {01},
Organization = {Cambridge Univ Press},
Pages = {106--109},
Volume = {48}
}

@Book{robert2004,
Title = {Monte Carlo statistical methods},
Author = {Robert, Christian P. and Casella, George},
Publisher = {Springer Texts in Statistics},
Year = {2004},
Edition = {second edition}
}

@Article{roberts2001,
author = {Roberts, Gareth O. and Rosenthal, Jeffrey S.},
journal = {Statistical Science},
number = {4},
pages = {351--367},
title = {Optimal scaling for various {M}etropolis-{H}astings algorithms},
volume = {16},
year = {2001}
}

@article{robins2007,
title={An introduction to exponential random graph (p*) models for social networks},
author={Robins, Garry and Pattison, Pip and Kalish, Yuval and Lusher, Dean},
journal={Social networks},
volume={29},
number={2},
pages={173--191},
year={2007},
publisher={Elsevier}
}

@InProceedings{stoehr-friel2015,
Title = {Calibration of conditional composite likelihood for Bayesian inference on Gibbs random fields},
Author = {Stoehr, Julien and Friel, Nial},
Booktitle = {JMLR W$\&$CP: Proceedings of the Eighteenth International Conference on Artificial Intelligence and Statistics},
Year = {2015},
Pages = {921--929},
Volume = {38}
}

@article{munro1951,
author = {Herbert Robbins and Sutton Monro},
journal = {The Annals of Mathematical Statistics},
number = {3},
pages = {400-407},
publisher = {Institute of Mathematical Statistics},
title = {A Stochastic Approximation Method},
volume = {22},
year = {1951}
}

@article{carpenterstan,
   author = {Bob Carpenter and Andrew Gelman and Matthew Hoffman and Daniel Lee and Ben Goodrich and Michael Betancourt and Marcus Brubaker and Jiqiang Guo and Peter Li and Allen Riddell},
   title = {Stan: A Probabilistic Programming Language},
   journal = {Journal of Statistical Software, Articles},
   volume = {76},
   number = {1},
   year = {2017},
   issn = {1548-7660},
   pages = {1--32},
   doi = {10.18637/jss.v076.i01}
}

@Article{tierney1998,
Title = {A note on {M}etropolis-{H}astings kernels for general state spaces},
Author = {Tierney, Luke},
Journal = {The Annals of Applied Probability},
Year = {1998},
Month = {02},
Number = {1},
Pages = {1--9},
Volume = {8},
Publisher = {The Institute of Mathematical Statistics}
}

@Article{bergm2014,
title = {{Bergm}: {B}ayesian Exponential Random Graphs in {R}},
author = {Alberto Caimo and Nial Friel},
journal = {Journal of Statistical Software},
year = {2014},
volume = {61},
number = {2},
pages = {1--25},
url = {http://www.jstatsoft.org/v61/i02/}
}

@Manual{giraf,
title = {GiRaF: A Toolbox for Gibbs Random Fields Analysis},
author = {Julien Stoehr and Pierre Pudlo and Nial Friel},
year = {2016},
note = {R package version 1.0},
url = {https://CRAN.R-project.org/package=GiRaF}
}

@Manual{cubature,
title = {cubature: Adaptive Multivariate Integration over Hypercubes},
author = {Balasubramanian Narasimhan and Steven G. Johnson},
year = {2017},
note = {R package version 1.3-8},
url = {https://CRAN.R-project.org/package=cubature},
}

@article{karateclub,
author = {Wayne W. Zachary},
journal = {Journal of Anthropological Research},
number = {4},
pages = {452--473},
publisher = {[University of New Mexico, University of Chicago Press]},
title = {An {I}nformation {F}low {M}odel for {C}onflict and {F}ission in {S}mall {G}roups},
volume = {33},
year = {1977}
}

@article{polyak1992,
title = {Acceleration of stochastic approximation by averaging},
author = {Polyak, Boris T and Juditsky, Anatoli B},
journal = {SIAM Journal on Control and Optimization},
volume = {30},
number = {4},
pages = {838--855},
year = {1992},
publisher = {SIAM}
}

@techreport{ruppert1988,
title = {Efficient estimations from a slowly convergent {R}obbins-{M}onro process},
author = {Ruppert, David},
year = {1988},
institution = {Cornell {U}niversity {O}perations {R}esearch and {I}ndustrial {E}ngineering},
number = {781}
}

@InProceedings{chen2014,
title = {Stochastic Gradient Hamiltonian Monte Carlo},
author = {Tianqi Chen and Emily Fox and Carlos Guestrin},
booktitle = {Proceedings of the 31st International Conference on Machine Learning},
pages = {1683--1691},
year = {2014},
editor = {Eric P. Xing and Tony Jebara},
volume = {32},
number = {2},
series = {Proceedings of Machine Learning Research},
address = {Bejing, China},
month = {22--24 Jun},
publisher = {JMLR.org}
}

@article{roberts1996,
author = {Roberts, Gareth O. and Tweedie, Richard L.},
journal = {Bernoulli},
month = {12},
number = {4},
pages = {341--363},
publisher = {Bernoulli Society for Mathematical Statistics and Probability},
title = {Exponential convergence of {L}angevin distributions and their discrete approximations"},
volume = {2},
year = {1996}
}

\end{document}